\documentclass[journal=jacsat,manuscript=article]{achemso}

\usepackage[version=3]{mhchem}
\usepackage{xcolor}

\author{Zhengxin Li}
\affiliation{Department of Chemical and Materials Engineering, University of Alberta, Alberta T6G 1H9, Canada}
\author{Hongbo Zeng}
\affiliation{Department of Chemical and Materials Engineering, University of Alberta, Alberta T6G 1H9, Canada}
\author{Xuehua Zhang}
\affiliation{Department of Chemical and Materials Engineering, University of Alberta, Alberta T6G 1H9, Canada}
\email{xuehua.zhang@ualberta.ca}

\title[An \textsf{achemso} demo]
  {Growth Rates of Hydrogen Microbubbles in Reacting Femtoliter Droplets}

\abbreviations{OTS, APTES, R6G}

\begin{document}

\begin{tocentry}
\centering
 \includegraphics[height=4.5cm]{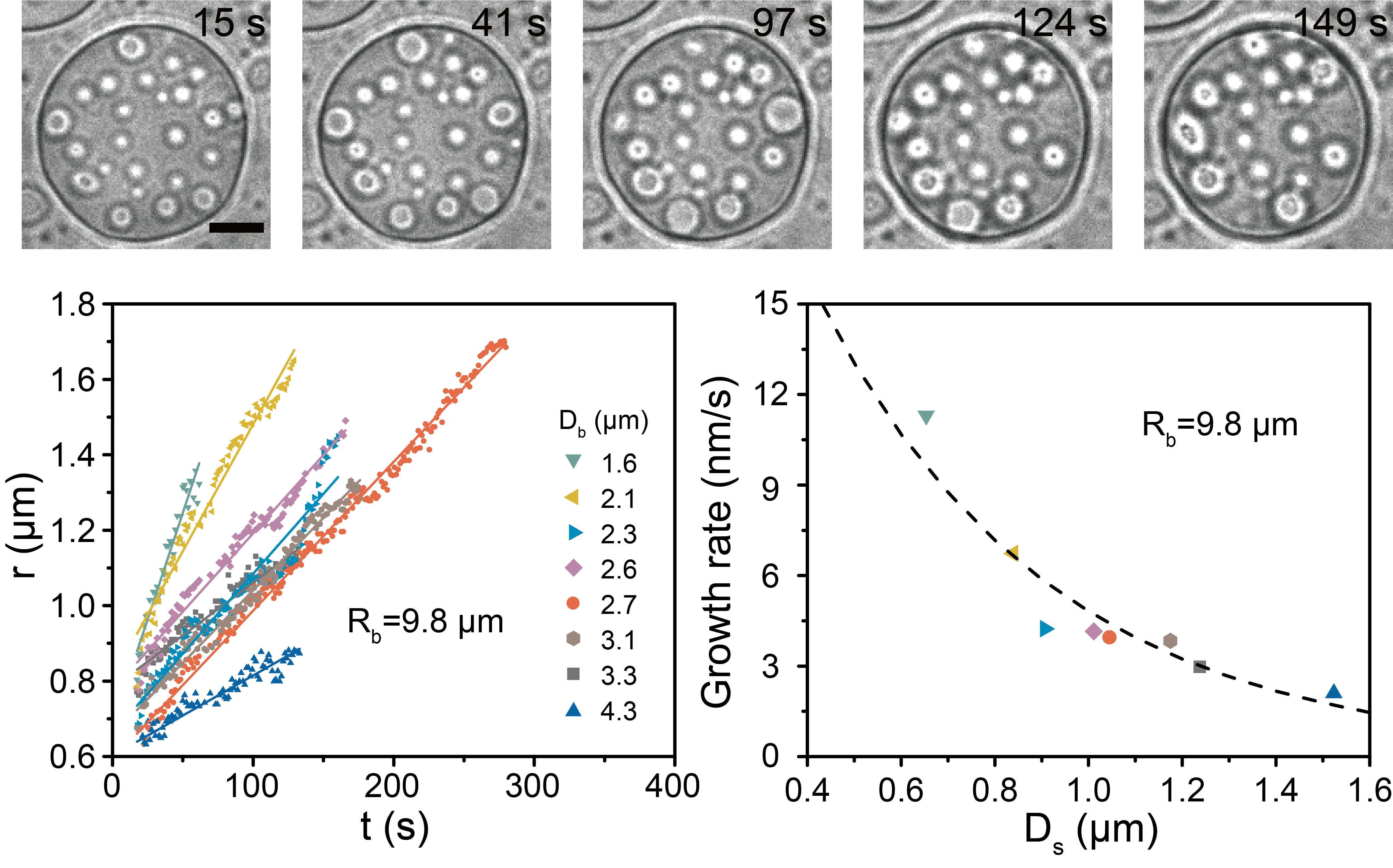}
\end{tocentry}

\begin{abstract}
Chemical reactions in small droplets are extensively explored to accelerate the discovery of new materials, increase efficiency and specificity in catalytic biphasic conversion and in high throughput analytic.
In this work, we investigate the local rate of gas-evolution reaction within femtoliter droplets immobilized on a solid surface.
The growth rate of hydrogen microbubbles ($\geq$ 500 nm in radius) produced from the reaction was measured online by high-resolution confocal microscopic images. 
The growth rate of bubbles was faster in smaller droplets, and of bubbles near the three-phase boundary in the same droplet.
The results were consistent for both pure and binary reacting droplets and on substrates of different wettability. 
Our theoretical analysis based on diffusion, chemical reaction and bubble growth in a steady state predicted that the concentration of the reactant diffusing from the surrounding depended on the droplet size and the bubble location inside the droplet, in good agreement with experimental results.
Our results reveal that the reaction rate may be spatially non-uniform in the reacting microdroplets.
The findings may have implications for formulating chemical properties and uses of these droplets.
\end{abstract}

\section{Introduction}
On-drop chemistry has been increasingly explored for fine chemical production based on biphasic reactions \cite{meng2019pickering}, synthesis of novel bio- and nano-materials \cite{kim2020plasmonic}, fast and sensitive chemical analysis \cite{wei2020integrated}, and engineering chemorobotic platform \cite{cronin2014,banno2021,hui2020}.
Notably, many reactions confined in the droplets have been reported to be faster than on a large scale.
The reaction rates may be enhanced by $10^2$ to even $10^6$ times compared to reactions in the bulk \cite{vannoy2021electrochemical,banerjee2015syntheses,bain2015accelerated}. Droplet reactions may also simplify post purification processes for a wide range of biphasic reactions where reactants or products present in two immersible fluid phases, such as oil and water or liquid and gas \cite{yang2015compartmentalization,zhang2016compartmentalized}. Microdroplet chemistry is considered to be green and sustainable,
enabling efficient chemical conversion for a wide range of reactants under mild reaction conditions or without using metal, heat, or expensive catalyst\cite{cheng2021}.

The fast chemical kinetics in small droplets is attributed to various interfacial phenomena \cite{kevin2020}, including molecular configuration, local concentration or partial solvation of reactants, or unusual rate of electron transfer at the gas-water interface \cite{nam2017abiotic,nam2018abiotic,nakatani1995direct,nakatani1995droplet,nakatani1996electrochemical}.
Electric potential energy localized at the interface may also alter the internal chemical equilibrium inside microdroplets \cite{chamberlayne2022microdroplets}. Recently \citet{zhong2020ultrafast} proposed that the surface of microdroplets may provide an energetically favourable environment for redox reactions in fast enzymatic protein digestion in microdroplet spray.  
In another case, reactant accumulation at the droplet surface was attributed to enhancement in the reaction between a droplet containing lipids and a droplet containing lipase \cite{burris2021enzyme}. The partial solvation of reactants at the interface may explain the accelerated reaction between microdroplets containing amines and $CO_2$ where the reaction was only confined at the droplet surface \cite{huang2021accelerated}. The acceleration factor increases with the decrease in the concentration of the reactant \cite{huang2021accelerated}.

Beyond significant acceleration in reaction rate, microdroplet reaction enables the spontaneous occurrence of reactions that is thermodynamically unfavourable. The type of reactions ranges from oxidation of water in air to biomolecules or origin of life in prebiotic earth, for instance, the generation of hydroperoxide \cite{lee2020condensing}, reduction of 2,6-dichlorophenolindophenol (DCIP) by ascorbic acid in absence of catalyst \cite{lee2015microdroplet}, and production of ribonucleotides from ribonucleosides \cite{ju2022aqueous}.

Higher chemical efficiency was reported not only for reacting droplets in the gas phase, but also for reactions between microdroplets and reactants dissolved in the surrounding liquid.
An example is Mannich reaction in emulsion droplets \cite{fallah2014enhanced}.
A comprehensive analytic model was developed by \citet{fallah2014enhanced} to explain the fast reaction rate of emulsion droplets.
The reaction-adsorption mechanism took into account the reaction equilibrium constant and forward rate constant associated with the concentration difference of the reactants throughout the droplet.  The mass flux between the droplets to the surrounding phase also plays an important role in the droplet reaction kinetics \cite{li2020speeding,li2021size}. In this regard, 
femtoliter droplets immobilized on a solid surface have been used as a model system for the quantitative study of droplet reaction rates. As the three-phase contact line of surface droplets is pinned by the solid surface, these droplets are stable on the substrate as the reactant is supplied in a controlled flow \cite{lohse2015surface,lei2015,levkin2018,lijuan2021}. The volume of surface droplets is conveniently controlled by the solvent exchange method \cite{zhang2015formation,qian2019,lei2016JPCL}. The enhanced gas production rate in smaller surface droplets has been revealed recently from the growth rate of hydrogen nanobubbles as the product \cite{dyett2020accelerated}. In particular, the bubble growth rates scaled with the droplet radius $R$ with a power law $R^{-n}$ with $n$ from 0.7 to 2.4.

As chemical acceleration is mainly attributed to the important impact from physical and chemical properties of the interface, the enhancement may be expected to decay to a certain extent with the distance away from the droplet surface. A remaining question is whether the reaction rate inside the droplets is uniform spatially.   In this work, we will focus on the local reaction rate inside femtoliter droplets. The growth rate of hydrogen bubbles from a gas-generating reaction in droplets will be followed by confocal microscopic imaging. 
In theoretical analysis of the local reaction rate, we take into account the reactant diffusion, chemical reaction equilibrium, and gas consumption by other bubbles coexisting in the droplet.
To the best of our knowledge, our experimental results provide direct evidence that the chemical kinetics and mass balance in the reacting microdroplets are not spatially uniform.
These findings may help us to better understand the biphasic reaction kinetics of gas evolution reaction of microdroplets and to design and control droplet reaction in nanomaterials fabrication, heterogeneous catalysis, and in-demand hydrogen bubble production.

\section{Experimental}
\subsection{Chemicals and materials}
Methylhydrosiloxane (Sigma Aldrich) was the reactive liquid in the droplets.
Octanol ($\geq$95\%, Fisher Scientific) was the non-reactive liquid in a binary droplet. 
Sodium hydroxide (NaOH) ($\geq$97\%, Alfa Aesar) acted as the catalyst for the gas-evolution reaction between siloxane droplets and water in the surrounding. 
All chemicals were used as received without any further purification.
Water (18.2 M$\Omega$cm) was purified by a Milli-Q purification unit (Millipore Sigma).

High precision cover glasses (60 mm length, 24 mm width, Azer Scientific) were hydrophobilized with 3-aminopropyl triethoxysilane (APTES) ($\geq$98\%, TCI America) and octadecyl trichlorosilane (OTS) ($\geq$95\% Fisher Scientific) by following protocols in the literature \cite{ramiasa2013contact,lessel2015,asenath2008prevent}.
Before use, hydrophobized glass substrates were sonicated in water and ethanol successively for 3 minutes and then dried in a stream of air.
The surface tension of the droplet liquid $\sigma_{liq-air}$ was tested with the drop shape analyzer (DSA-100, Kruss).
The contact angle $\theta$ of surface microdroplets on different substrates were constructed from 3D images collected from a scanning laser confocal microscope (Leica Stellaris 5) through 20$\times$ objectives (0.60 NA).
Table \ref{tbl:Property} and \ref{tbl:CA} list relevant physical properties of droplet liquids and contact angle of surface droplets $\theta$ on various types of the substrates.
Droplet formation and chemical reactions were conducted at the room temperature of $\sim$ 21$^\circ$C.

\begin{table}[htbp]
\small
  \caption{\ Physical properties of liquids used in droplet formation}
  \label{tbl:Property}
  \begin{tabular*}{0.4\textwidth}{@{\extracolsep{\fill}}ccc}
    \hline
    Liquid & $\sigma_{liq-air}$ & Viscosity $\mu$ \\
    (vol) & (mNm$^{-1}$) & (mPa$\cdot$s) \\
    \hline
    Water & 69 &  1\\
    Siloxane & 19 & 15 $\sim$ 30 \\
    Octanol & 21 & 9 \\
    \hline
  \end{tabular*}
\end{table}

\begin{table}[htbp]
\small
  \caption{Contact angle of femtoliter droplets on the substrates used in our experiments}
  \label{tbl:CA}
  \begin{tabular*}{0.5\textwidth}{@{\extracolsep{\fill}}ccc}
    \hline
    Droplet liquid & Substrate & $\theta$ ($^\circ$) \\
    \hline
    Siloxane & 1 & $\sim$ 26($\pm$6) \\
    Siloxane & 2 & $\sim$ 31($\pm$5) \\
    Siloxane & 3 & $\sim$ 68($\pm$9) \\
    Siloxane + Octanol & 3 & $\sim$ 32($\pm$7) \\
    Siloxane + Octanol & 4 & $\sim$ 23($\pm$5) \\
    \hline
  \end{tabular*}
\end{table}

\subsection{Formation of surface microdroplets of reactive liquid}
\begin{figure}[htbp]
\centering
 \includegraphics[height=9cm]{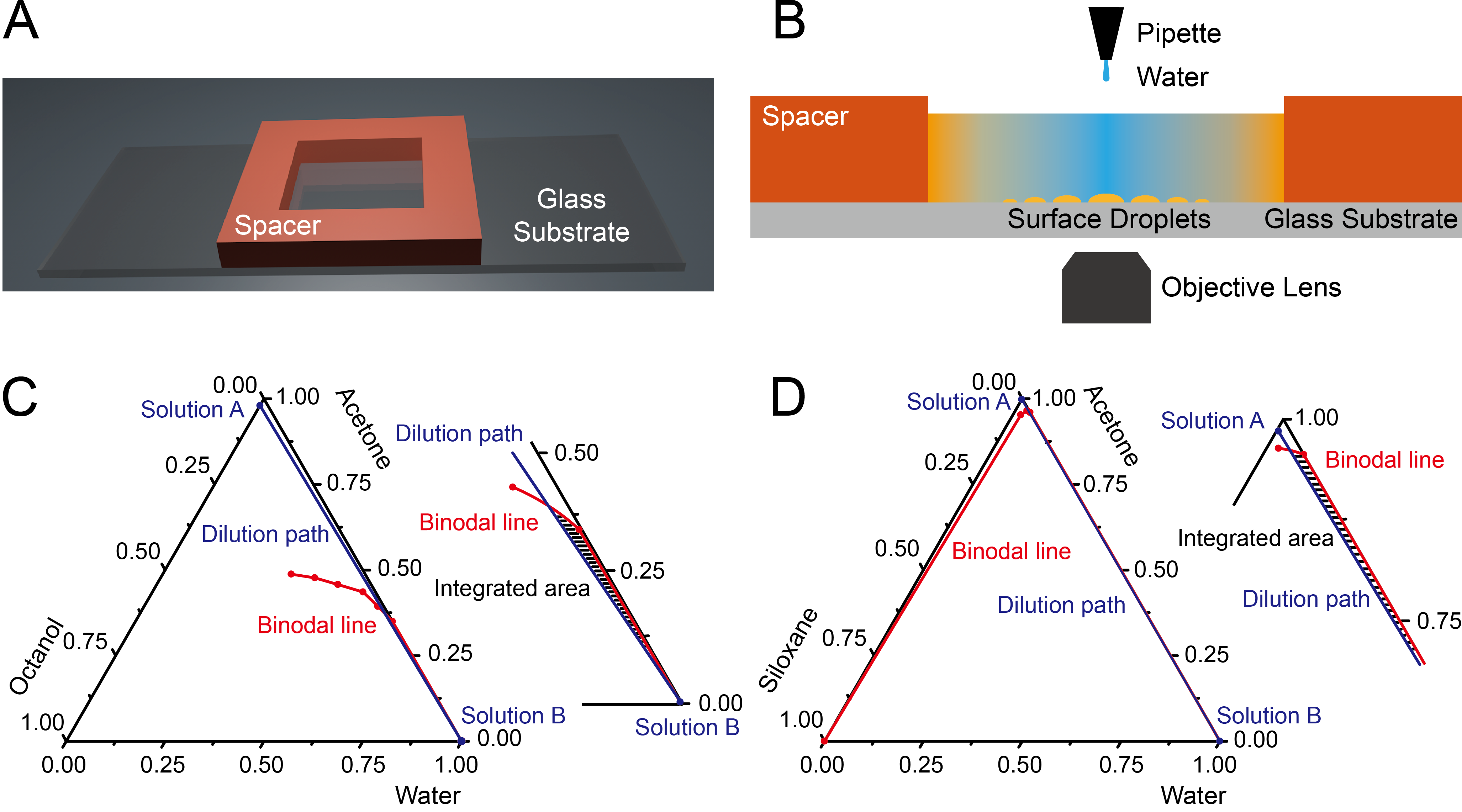}
  \caption{(A) The sketch of the well reactor used in our experiments for formation and reaction of droplets.
  (B) Illustration of preparing surface microdroplets by solvent exchange.
  (C)\&(D) Ternary phase diagrams of (A) Siloxane-acetone-water system and (B) Octanol-acetone-water system.
  Points labelled as solution A\&B are respectively the composition points of solution A\&B.
  Black shaded areas surrounded by binodal lines (red) and dilution paths (blue) represent the oversaturation of each component in the mixing front.}
\label{fgr:Setup1}
\end{figure}

Surface microdroplets were prepared within a well reactor as sketched in Figure \ref{fgr:Setup1}A.
The well reactor consisted of a piece of the glass substrate and a silicone rubber spacer.
The well was 14 mm in both length and width, and 3 mm in depth.
Surface microdroplets were formed by the standard solvent exchange process where the solution of droplet liquid (Solution A) was displaced by a poor solvent for droplet liquid (Solution B). 
In our experiments, Solution A was 1 vol\% siloxane in acetone ($\geq$99.5\%, Fisher Scientific). Solution B was water. 

During the solvent exchange, the well was initially filled with 200 $\mu$l of Solution A.
Then a micropipette was used to add 200 $\mu$l of solution B into the well at a rate of 2 drops per 10 s.
Then 200 $\mu$l of the liquid was removed from the well by a micropipette at the same rate.
Addition of Solution B into the well-induced formation and growth of surface microdroplets on the substrate, as sketched in Figure \ref{fgr:Setup1}B.
The process of addition-removal of solution B was repeated 4 times till the liquid in the well became clear and surface microdroplets formed on the wall. At the completion of the solvent exchange, the liquid in the well was not pure water, but an aqueous solution with a volume of  400 $\mu$l contained acetone of $>$ 6\%, v/v.
The presence of acetone in the surrounding phase was determined by a microvolume ultraviolet-visible (UV-Vis) spectrophotometer (Nanodrop 2000c, Thermo Fisher).
Figure S1 in supporting materials demonstrates the UV-Vis absorption spectrum of a sample from the surrounding phase with $\sim$ 9\% (v/v) acetone.

\subsection{Formation of surface microdroplets of reactive and non-reactive liquids}
Binary droplets consisting of octanol and siloxane were prepared on both APTES-coated and OTS-coated substrates.
Solution A was 2 vol\% octanol and 0.3 vol\% siloxane in acetone solution. 
A trace amount of fluorescence dye rhodamine 6G (R6G, $\sim$ 5 $\mu$M) was added into Solution A for visualization by confocal imaging.
Solution B was water.
The solvent exchange was performed in the well reactor by following the same procedure as above to form binary droplets on the substrate. By the end of the solvent exchange, the concentration of acetone in the solution inside the well was around 6\% (v/v), and the volume of the solution was controlled to be at 400 $\mu$l.

The composition of binary droplets was predetermined by the oversaturation level of octanol and siloxane during the solvent exchange \cite{li2018formation}, which could be approximately estimated according to the difference of dilution path and the binodal curve in the solubility phase diagram. The detailed protocol and the analysis were reported in literature \cite{lu2016,li2018formation}.
Ternary phase diagrams of octanol-acetone-water and siloxane-acetone-water were prepared by titration and demonstrated in Figure \ref{fgr:Setup1}C\&D.
The area surrounded by binodal curves and dilution path in the ternary phase diagram represents the overall oversaturation (octanol:siloxane $\sim$ 4:1).

\subsection{Tracking microbubbles in reacting droplets}
50 $\mu$l of 0.24 M NaOH aqueous solution was added into the well by a micropipette to trigger the reaction between siloxane and water.
The existence of acetone was confirmed to be necessary for the bubble formation by controlled experiments, in which acetone was almost removed by repeating adding water and removing the mixture from the well 12 times.

Laser scanning confocal microscopes (Leica SP5 and Leica Stellaris 5) through 100$\times$ objectives (1.44 NA/1.49 NA) were employed to track the reaction process in-situ.
Bright-field images and confocal images of the bubble formation process were respectively recorded with a transmission detector and a hybrid detector.
488/534 nm laser beams were used to excite the dye R6G incorporated in droplets.
The pixel sizes of the videos range from 151 nm to 303 nm.
The frame rate was 0.77 fps.

Open source PIMS, scikit, and trackPy package for python, combined with ImageJ, were applied to analyze the images.
Droplets and bubbles in the field of view were processed frame by frame by home-built python code.
The base areas of hydrogen bubble $a_b$ and reacting droplet $A_b$ were extracted as a function of time $t$.
Time zero ($t_0$) was defined as the moment when NaOH solution was added to the well.
Distance from the center of the bubble base to the three-phase contact line of the droplet $D_b$ was measured by ImageJ.
In most experiments, droplets and bubbles on our homogeneous substrates were assumed to keep the shape of a spherical cap without the strong pinning effect.
Some bubbles in small droplets ($R_b<$ 5 $\mu$m) grew irregularly and were not considered when calculating the average growth rate of hydrogen bubbles.

\section{Results and discussion}
\subsection{Theoretical model: Local reaction rate in microdroplets}
\begin{figure}[htbp]
\centering
 \includegraphics[height=4.6cm]{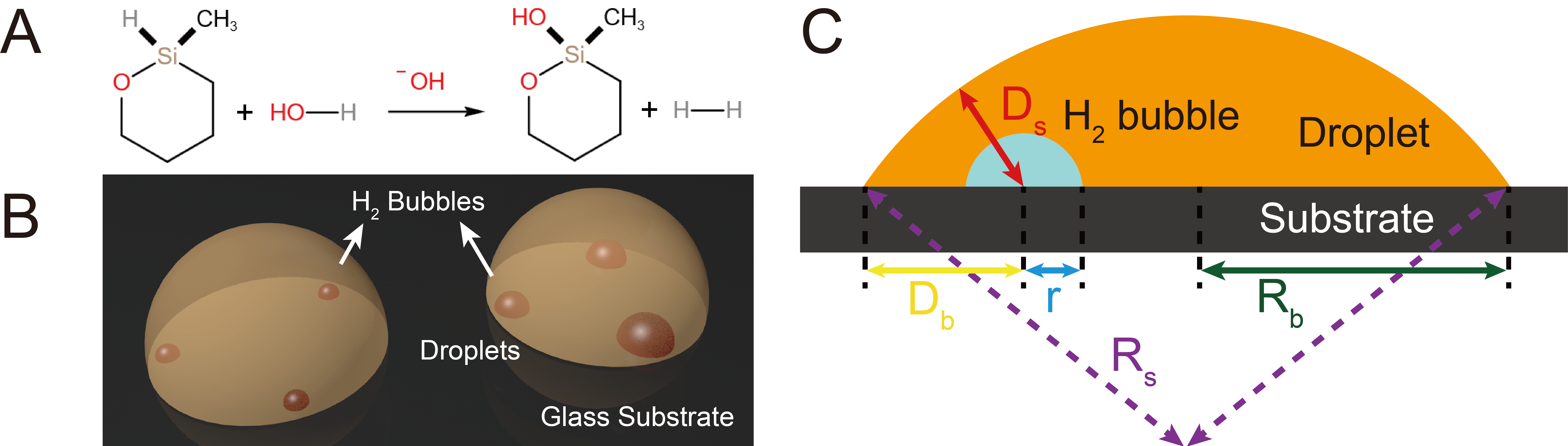}
  \caption{(A) The chemical equation of siloxane dehydrocoupling with water.
  (B) Illustration of hydrogen bubbles formed in droplets.
  (C) Base radius of droplet $R_b$ (green) and bubble $r$ (blue), radius of the curvature $R_s$, distance from the bubble center to the droplet rim $D_b$ (yellow) and the droplet surface $D_s$ (red) in the sketch of a surface microdroplet.}
\label{fgr:Setup2}
\end{figure}

The chemical reaction of siloxane dehydrocoupling is catalyzed by hydroxide from the aqueous phase (Figure \ref{fgr:Setup2}).
At first, hydroxide in bulk attacks silicon atoms of siloxane.
Then, hydroxide is reformed by consuming water, and hydrogen is liberated.
The oversaturation of the hydrogen product inside the reacting droplet leads to the formation and growth of multiple hydrogen bubbles. 

The dimensions of a bubble and a reacting droplet are sketched in Figure \ref{fgr:Setup2}B.
Base radius of bubble $r$ and droplet $R_b$ were calculated from the base area of bubble $a_b$ and droplet $A_b$ as $r=(a_b/\pi)^{0.5}$ and $R_b=(A_b/\pi)^{0.5}$.
The radius of curvature $R_s$ of the droplet was calculated as $R_s=R_b/sin\theta$.
Based on the contact angle of the droplet $\theta$, the base radius $R_b$, the curvature radius $R_s$, and the distance from the bubble center to the droplet rim $D_b$, the shortest distance from the bubble center to the droplet surface $D_s$ can also be calculated by the following equation:
\begin{equation}
\label{eq:distance}
  D_s = R_s-\sqrt{(R_b-D_b)^2+R_b^2/tan^2\theta}.
\end{equation}

The reaction in our experiments is irreversible.
The rate-limiting step in local gas production in the droplet is water (the reactant) diffusion into the droplets, a step that determines the location and the rate of the bubbles.
Hydrogen produced from the reaction located at the droplet will be discussed in a later section.
The reaction with siloxane consumed water and reduced the water concentration in the droplet.
The mass balance of water in the droplet consists of the water diffusion and the chemical reaction.
In a spherical droplet, the water concentration profile will be analyzed in the radial direction.
The mass balances of water inside the droplet can be given by Fick's law:
\begin{equation}
\label{eq:concentration1}
  \partial_tC_w(D_s) = D\nabla^2 C_w(D_s)-k_rC_w(D_s)C_{SiH}.
\end{equation}
$\partial_t$ is the time derivative.
$\nabla$ is the Nabla operator.
$D$ is the diffusion coefficient of the hydrogen in and out of the droplet.
$C_w(D_s)$ is the local concentration of water with the distance $D_s$ from the droplet surface.
$C_{SiH}$ is the concentrations of siloxane, which can be taken as a constant in the droplet.
$k_r$ is the rate constant of the reaction.
At the point with distance $D_s$ from the droplet surface, the variation of water concentration $\partial_tC_w(D_s)$ is given by the water diffusion $D\nabla^2 C_w(D_s)$ minus the water consumption by the chemical reaction $k_rC_w(D_s)C_{SiH}$.

Assuming the steady state that water concentration inside the droplet was constant with time, $\partial_tC_w(D_s)$ was zero and can be neglected in equation \ref{eq:concentration1}:
\begin{equation}
\label{eq:concentration2}
  k_rC_w(D_s)C_{SiH} = D\nabla^2 C_w(D_s).
\end{equation}
We defined a characteristic diffusion length $\epsilon$ as $\sqrt{D/(k_rC_{SiH}})$ to describe the competing effect from water diffusion from the droplet surface and from water consumption by the chemical reaction in the droplet \cite{fallah2014enhanced}.
Solving equation \ref{eq:concentration2} (see Supporting Materials for the derivation), we get:
\begin{equation}
\label{eq:concentration3}
  C_w(D_s) = C_w(D_s)|_{D_s=0}\frac{R_s\sinh{(R_s-D_s)/\epsilon}}{(R_s-D_s)\sinh{R_s/\epsilon}}.
\end{equation}
Based on equation \ref{eq:concentration3}, water concentration at a distance $D_s$ to the droplet surface $C_w(D_s)$ is determined by the water concentration nearing the droplet surface $C_w(D_s)|_{D_s=0}$, radius of the droplet curvature $R_s$, and diffusion length $\epsilon$.
From the profile of $C_w(D_s)$ in equation \ref{eq:concentration3}, the profile of production rate of hydrogen $\dot m(D_s)$ throughout the droplet can be given by $\dot m(D_s)=k_rC_{SiH}C_w(D_s)$, as:
\begin{equation}
\label{eq:concentration4}
\begin{aligned}
  \dot m(D_s) = k_rC_{SiH}C_w(D_s)|_{D_s=0}\frac{R_s\sinh{(R_s-D_s)/\epsilon}}{(R_s-D_s)\sinh{R_s/\epsilon}}\\
  = \dot m(D_s)|_{D_s=0}\frac{R_s\sinh{(R_s-D_s)/\epsilon}}{(R_s-D_s)\sinh{R_s/\epsilon}}.
  \end{aligned}
\end{equation}
$\dot m(D_s)|_{D_s=0}$ is the theoretical hydrogen production rate nearing the droplet surface.

In the above analysis, the profile of hydrogen production rate is obtained along the radial direction inside a spherical droplet.
In surface droplets with contact angles much lower than 90$^{\circ}$, microbubbles on the substrate are close to the droplet surface than to the center of the sphere.
Hence the radial concentration profile from equation \ref{eq:concentration4} services a good approximation for the hydrogen supply for microbubbles over the base area of the droplet.
Assuming a balanced state that the hydrogen product surrounding the bubble diffuses to the bubble surface and is consumed by the bubble growth, the average bubble growth rate $\dot r(D_s)$ should be approximately proportional to $\dot m(D_s)$ in equation \ref{eq:concentration4} and can be fit by the same profile:
\begin{equation}
\label{eq:concentration5}
  \dot r(D_s) = \dot r(D_s)|_{D_s=0}\frac{R_s\sinh{(R_s-D_s)/\epsilon}}{(R_s-D_s)\sinh{R_s/\epsilon}}.
\end{equation}
$\dot r(D_s)|_{D_s=0}$ is the theoretical average growth rate of the hydrogen bubble nearing the droplet surface.

\subsection{Growing microbubbles in reacting surface droplet}
\begin{figure}[htbp]
\centering
 \includegraphics[height=4.5cm]{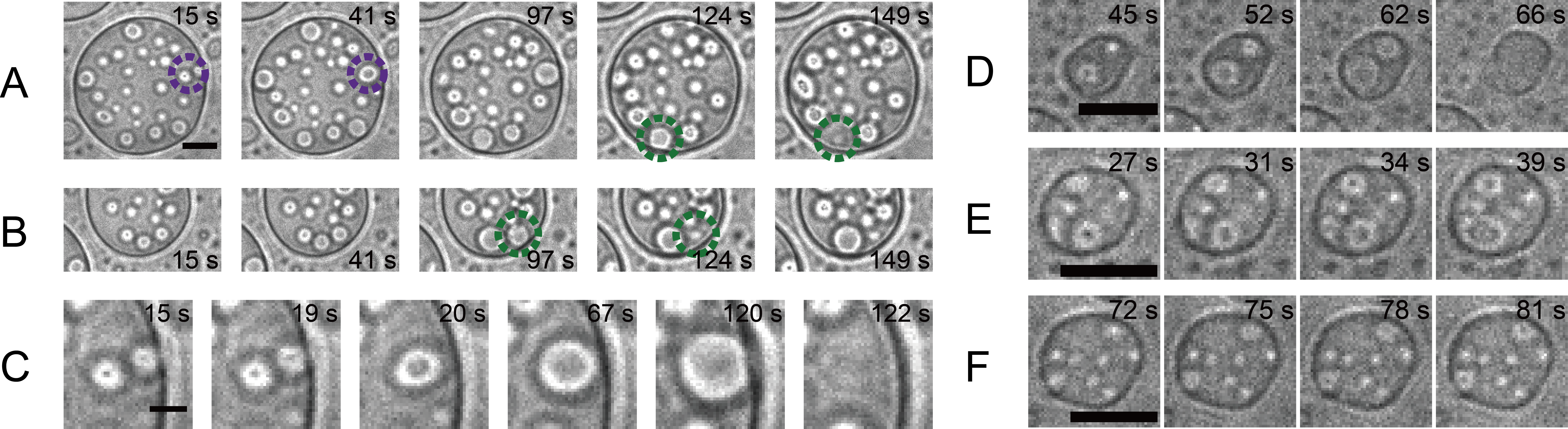}
  \caption{(A)\&(B) Screenshots of hydrogen bubbles in siloxane droplets with the base radius of 9.8 $\mu$m (A) and 7.5 $\mu$m (B).
  Purple circles and green circles denote where bubbles coalesce and collapse.
  (C) Zoom-in of two bubbles in (A) from coalescence to detachment.
  (D)-(E) Screenshots of hydrogen bubbles in droplets with the base radius of 1.8 $\mu$m (A), 2.2 $\mu$m (B), and 2.8 $\mu$m (C).
  The length of scale bars in (A)(B)(D)(E) are 5 $\mu$m, and in (C) is 2 $\mu$m.}
\label{fgr:A1}
\end{figure}

The contact angle of the reacting droplets constructed from 3D confocal images were respectively $\sim$ 26$^\circ(\pm6^\circ)$ (Figure \ref{fgr:A1}A\&B) and $\sim$ 31$^\circ(\pm5^\circ)$ (Figure \ref{fgr:A1}D-F).
After introducing NaOH solution into the reactor, microbubbles nucleate throughout reacting droplets.
The radius of microbubbles ranges from a few hundred nanometers to several microns.
Screenshots in Figure \ref{fgr:A1}A\&B show the continuous growth of bubbles inside two large droplets with the base radius $R_b$ of 9.8 $\mu$m and 7.5 $\mu$m.
Figure \ref{fgr:A1}D-E show hydrogen bubbles grow in small droplets with $R_b$ less than 4 $\mu$m.
The number of bubbles is much higher in a larger droplet.

As bubbles grew with time, more of the droplet base area was taken up by bubbles.
Two adjacent bubbles may coalesce and merge into a larger bubble.
The elliptical morphology of bubbles after coalescence suggests that bubbles are pinned by the substrate due to the slow relaxation in the viscose droplet.
For large bubbles with the base radius $r$ larger than 2 $\mu$m near the droplet rim, they may collapse and detach from the substrate.
Figure \ref{fgr:A1}C enlarges two representative bubbles in Figure \ref{fgr:A1}A from coalescence to detachment.
The collapse of the bubble near the rim may be due to the rapture of the oil thin film between the bubble and the aqueous phase outside the droplet \cite{dyett2020accelerated}.
More quantitative study of bubble coalescence and bubble detachment on the solid surface was reported in previous work \cite{PhysRevLett.127.235501}.

\begin{figure}[htbp]
\centering
 \includegraphics[height=9.1cm]{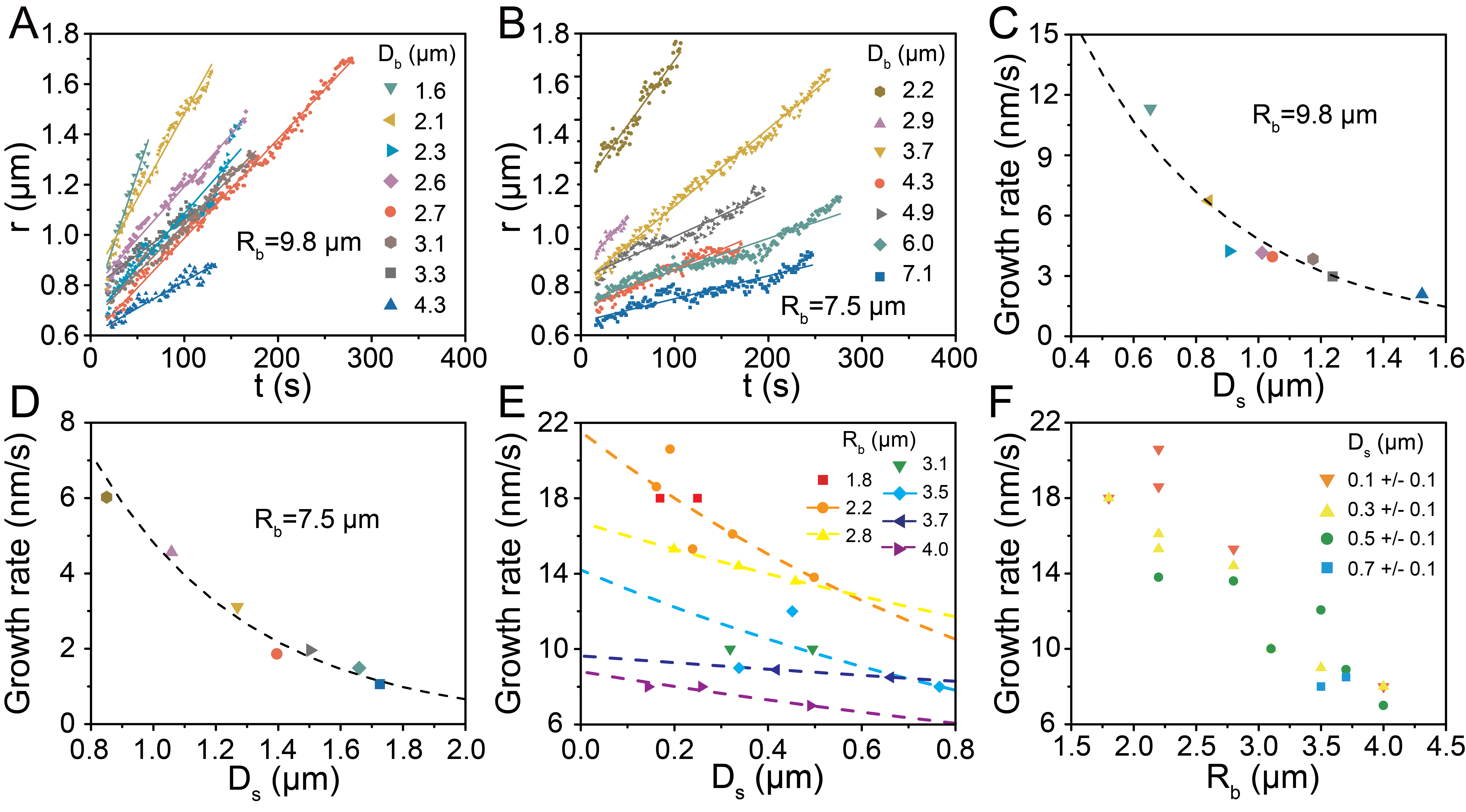}
  \caption{(A)\&(B) Base radius $r$ of hydrogen bubbles from Figure \ref{fgr:A1}A\&B as a function of time. 
  (C)\&(D) The average growth rate of hydrogen bubbles in (A)\&(B) as a function of the distance from the droplet surface to the bubble center $D_s$.
  Dashed lines in (C)-(E) are from the fitting of equation \ref{eq:concentration5}.
  (E) The average growth rate of hydrogen bubbles as a function of the distance from the bubble to the droplet surface $D_s$.
  Bubbles in (E)\&(F) are from droplets with the base radius $R_b$ ranging from 1.8 $\mu$m to 4.0 $\mu$m.
  (F) The average growth rate of hydrogen bubbles as a function of the base radius of the droplet $R_b$.}
\label{fgr:A2}
\end{figure}

For multiple bubbles in a droplet, bubbles around the droplet rim grew faster than those near the droplet center.
For example, from $t=$15 s to $t=$97 s in Figure \ref{fgr:A1}A\&B, bubbles at the droplet rim became $\sim$ 40\% larger in base radius, while bubbles far from the droplet rim remained almost the same size.
In Figure \ref{fgr:A2}A\&B, the base radius of bubbles $r$ was plotted as the function of $t$.
We approximate that bubbles grew linearly and calculate the average growth rate of each bubble by linear fitting. 
The distance from the bubble center to the droplet surface $D_s$ was also calculated by equation \ref{eq:distance}.
Figure \ref{fgr:A2}C\&D show that the bubble growth rate decreases with the increase in $D_s$. 
The experimental data can be well fitted with the theoretic model.

On the APTES-Si substrate where the contact angle of droplets was $\sim$ 31$^\circ(\pm5^\circ)$, bubbles in seven small droplets with $R_b$ ranging from 1.8 $\mu$m to 4.0 $\mu$m were followed with time.
The average growth rates of bubbles were approximately calculated by linear fitting, and plotted as a function of $D_s$ in Figure \ref{fgr:A2}E.
Overall the influence from $D_s$ in small droplets was relatively trivial.
Figure \ref{fgr:A2}F shows that the growth rate of bubbles increased with the decrease in the droplet size.

Table \ref{tbl:Parameter1} summarize the parameters $\epsilon$ and $\dot r(D_s)|_{D_s=0}$ in the theoretical fittings plotted in Figure \ref{fgr:A2}C-E.
$\dot r(D_s)|_{D_s=0}$ is higher in smaller droplets, consistent with the faster growth rate of bubbles in smaller droplets in the same group of experiments.
The enhanced reaction rate in smaller droplets was also found and quantitatively analyzed in our previous results  \cite{dyett2020accelerated,li2021size}. 
The accelerated bubble kinetics in smaller droplets was attributed to the faster accumulation of the hydrogen product in smaller droplets with a higher surface-to-volume ratio.

\begin{table}[htbp]
\small
  \caption{Fitting parameters $\epsilon$ and $\dot r(D_s)|_{D_s=0}$ for experimental results from siloxane droplets in Figure \ref{fgr:A2}.
  $R_b$ and $R_s$ are the base radius and curvature radius of the droplet.
  $\theta$ is the contact angle of the droplet.
  $D$, $k_r$, and $C_{SiH}$ are the hydrogen diffusion coefficient in the droplet, rate constant of the reaction, and siloxane concentration in the droplet.}
  \label{tbl:Parameter1}
  \begin{tabular*}{0.9\textwidth}{@{\extracolsep{\fill}}cccc}
    \hline
    $R_b$/$R_s$ ($\mu$m) & $\theta$ ($^\circ$) & $\epsilon=\sqrt{D/(k_rC_{SiH})}$ ($\mu$m) & $\dot r(D_s)|_{D_s=0}$ (nm/s) \\
    \hline
    9.8/22.4 & $\sim$ 26$^\circ\pm6^\circ$ & 0.5 & 34 \\
    7.5/17.1 & $\sim$ 26$^\circ\pm6^\circ$ & 0.5 & 36 \\
    4.0/6.1 & $\sim$ 31$^\circ\pm5^\circ$ & 1.7 & 9 \\
    3.7/5.6 & $\sim$ 31$^\circ\pm5^\circ$ & 3.4 & 10 \\
    3.5/5.3 & $\sim$ 31$^\circ\pm5^\circ$ & 1.2 & 14 \\
    2.8/4.3 & $\sim$ 31$^\circ\pm5^\circ$ & 1.8 & 17 \\
    2.2/3.4 & $\sim$ 31$^\circ\pm5^\circ$ & 1.0 & 22 \\
    \hline
  \end{tabular*}
\end{table}

\subsection{Effects from the contact angle of the reacting droplet}
\begin{figure}[htbp]
\centering
 \includegraphics[height=7cm]{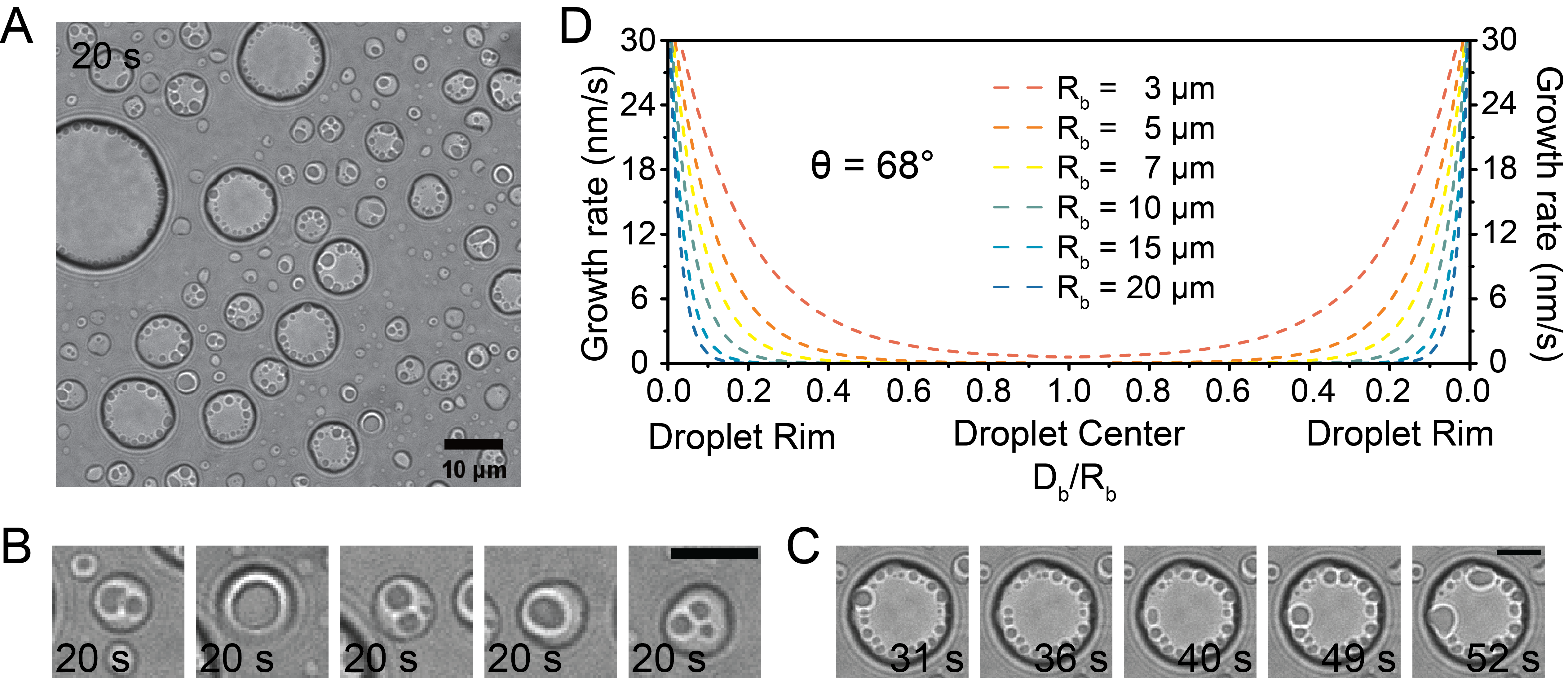}
  \caption{(A) Hydrogen bubbles in siloxane droplets with the contact angle of 68$^\circ(\pm9^\circ)$.
  (B) Zoom-in of small droplets ($R_b<$ 3 $\mu$m) from (A).
  (C) Screenshots of hydrogen bubbles grew in a siloxane droplet with the base radius of 12.3 $\mu$m.
  Scale bars in (B)\&(C) were 5 $\mu$m.
  (D) Theoretical average growth rate profile of hydrogen bubbles crossing the droplet base.
  The contact angle $\theta$ was taken as 68 $^\circ$.
  $\epsilon$ and $\dot m(D_s)|_{D_s=0}$ were taken as 0.5 $\mu$m and 34 nm/s, same the fitting parameters in Figure \ref{fgr:A2}C.}
\label{fgr:Ring}
\end{figure}

Both surface coating and the liquid composition in the droplet were varied to provide three different contact angles in our experiments from 23$^\circ$ to 68$^\circ$.
On an APTES-Si substrate where siloxane droplets were produced with the contact angle of $\sim$ 68$^\circ(\pm9^\circ)$, bubbles were found to exclusively nucleate at the droplet rim.
As shown in Figure \ref{fgr:Ring}A, a necklace of microbubbles developed around the rim of the droplet in contact with the basic solution.
For small droplets with the base radius $R_b$ less than 2 $\mu$m, as in Figure \ref{fgr:Ring}B, only one or two bubbles in most cases can be accommodated in a droplet.
Screenshots in Figure \ref{fgr:Ring}C demonstrates the evolution of bubbles in a representative droplet with $R_b=$ of 12.3 $\mu$m.

From equations \ref{eq:distance}\&\ref{eq:concentration5}, $\dot r(D_s)$ was approximately estimated as a function of normalized distance to the droplet rim $D_b/R_b$ crossing the droplet base, as shown in Figure \ref{fgr:Ring}D.
The contact angle of droplets was taken as $68^\circ$.
$\epsilon$ and $\dot r(D_s)|_{D_s=0}$ were set as 0.5 $\mu$m and 34 nm/s, same as the fitting parameters in Figure \ref{fgr:A2}C.
Apart from droplet with $R_b$ of 3 $\mu$m, $\dot r(D_s)$ rapidly decease below 1 nm/s before reaching the droplet center.
Taking a droplet with the base radius $R_b$ of 10 $\mu$m as the example, $\dot r(D_s)$ already decreased to 0.14 nm/s at $D_b/R_b$ of 0.3, suggesting an extremely low water concentration and slow reaction rate at $D_b=$ 3 $\mu$m.
It may take a long time for hydrogen to reach the concentration level for bubbles to nucleate at such a slow reaction rate.
The rapid decrease in water concentration inside droplets with high contact angles explains why bubbles only nucleate and grow at the droplet rim.

\begin{figure}[htbp]
\centering
 \includegraphics[height=9.1cm]{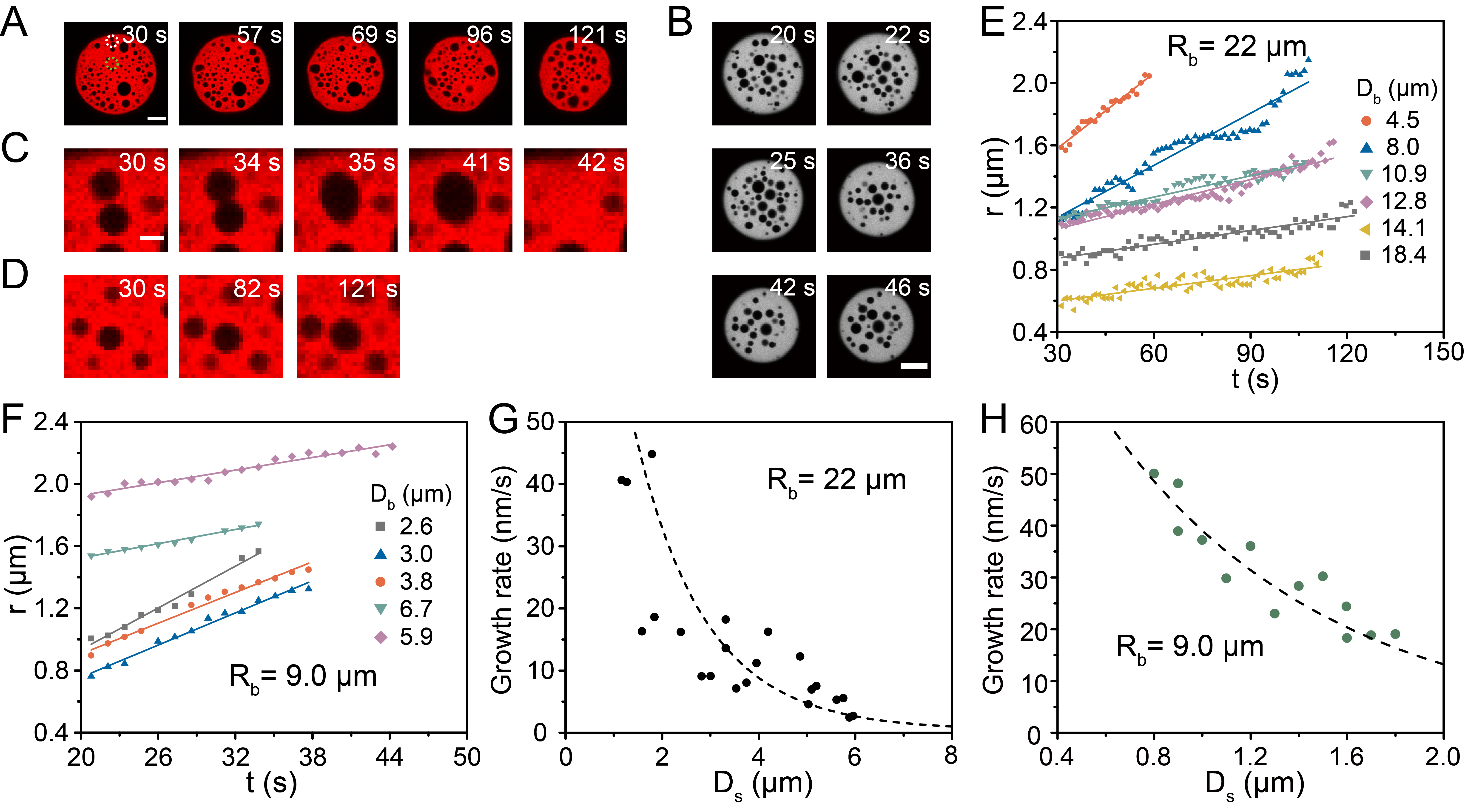}
  \caption{(A)\&(B) Bubbles grow in a binary droplet on an APTES-Si substrate (A) and an OTS-Si substrate (B).
  The base radius of the droplet $R_b$ are 22 $\mu$m in (A) and 9.0 $\mu$m in (B).
  The contact angle of binary droplets on the substrate were (A) $\sim$ 32$^\circ(\pm7^\circ)$ and (B) $\sim$ 23$^\circ(\pm5^\circ)$.
  The scale bars are (A) 10 $\mu$m and (B) 5 $\mu$m.
  (C)\&(D) Zoom-in images of hydrogen bubbles at the droplet rim (C) and the droplet center (D).
  (C)\&(D) are taken from the white and green circle in (A).
  The scale bar for (C)\&(D) in (C) was 2 $\mu$m.
  (E)\&(F) Base radius r of several representative hydrogen bubbles from (A)\&(B) as a function of time.
  (G)\&(H) Average growth rate of hydrogen bubbles from (A)\&(B) as a function of distance to the droplet surface $D_s$.
  Dashed lines in (G)\&(H) are from the fitting of equation \ref{eq:concentration5}.}
\label{fgr:B1}
\end{figure}

Screenshots in Figure \ref{fgr:B1}A\&B demonstrate the bubble growth in binary droplets on two substrates.
The contact angle of binary droplets were 32$^\circ(\pm7^\circ)$ on the OTS-Si substrates and $\sim$ 23$^\circ(\pm5^\circ)$ on the APTES-Si substrates.
$R_b$ of the binary droplets were 22 $\mu$m and 9.0 $\mu$m.
Similar to bubbles in pure droplets shown in Figure \ref{fgr:A1}, bubbles nucleated and grew throughout binary droplets.
Figure \ref{fgr:B1}C\&D respectively demonstrate hydrogen bubbles close to the droplet rim and at the droplet center.
The lifetime for a bubble at the droplet rim is usually less than 10 seconds from nucleation to detachment, much shorter than that of inner bubbles.
Figure \ref{fgr:B1}E\&F present sizes of several representative bubbles with different $D_b$ to the droplet rim.
Bubbles at the droplet rim grow much faster than bubbles at the droplet center.

Average growth rates of bubbles are plotted as a function of $D_s$ in Figure \ref{fgr:B1}G\&H.
Although data points were scattered in binary droplets, the overall trend was that there was a strong dependence of growth rate on $D_s$.
On the other hand, even the siloxane concentration in binary droplets was lower, the growth rate of bubbles near the droplet rim can go up to $\sim$ 50 nm/s, much faster than bubble growth in pure siloxane droplets.
As a result, both bubble coalescence and bubble detachment were more frequent in binary droplets than in pure droplets.

Dashed lines in Figure \ref{fgr:B1}G\&H are the fittings of experimental results by our theoretic model.
Table \ref{tbl:Parameter2} summarize the fitting parameters $\epsilon$ and $\dot r(D_s)|_{D_s=0}$ of binary droplets.
Corresponding to experimental results that bubble kinetics in binary droplets was faster than in pure droplets, $\dot r(D_s)|_{D_s=0}$ was also found to be higher than that of pure droplets.
The accelerated bubble kinetics in binary droplets was probably from the increased water solubility with the addition of polar octanol into the droplets.
Higher water concentration in the droplet contributes to a faster chemical reaction rate.

\begin{table}[htbp]
\small
  \caption{Fitting parameters $\epsilon$ and $\dot r(D_s)|_{D_s=0}$ for experimental results from octanol-siloxane binary droplets.
  $R_b$ and $R_s$ are the base radius and curvature radius of the droplet.
  $\theta$ is the contact angle of the droplet.
  $D$, $k_r$, and $C_{SiH}$ are the hydrogen diffusion coefficient in the droplet, rate constant of the reaction, and siloxane concentration in the droplet.}
  \label{tbl:Parameter2}
  \begin{tabular*}{0.9\textwidth}{@{\extracolsep{\fill}}cccc}
    \hline
    $R_b$/$R_s$ ($\mu$m) & $\theta$ ($^\circ$) & $\epsilon=\sqrt{D/(k_rC_{SiH})}$ ($\mu$m) & $\dot r(D_s)|_{D_s=0}$ (nm/s) \\
    \hline
    22.0/41.5 & $\sim$ 32$^\circ\pm7^\circ$ & 1.2 & 143 \\
    9.0/23.0 & $\sim$ 23$^\circ\pm5^\circ$ & 0.8 & 137 \\
    7.4/18.9 & $\sim$ 23$^\circ\pm5^\circ$ & 0.9 & 154 \\
    \hline
  \end{tabular*}
\end{table}

\subsection{Further discussion: On-droplet and in-droplet hydrogen production}
The reaction in our experiments can certainly take place on the surface of the droplets, due to abundant siloxane, water and the catalyst in the surrounding.
On-droplet hydrogen is expected to be limited in amount.  Moreover, the molecules at the interface may react even more readily to react \cite{vannoy2021electrochemical,banerjee2015syntheses,bain2015accelerated,kevin2020}.
Many reactions that almost do not happen when reactants are located in immiscible phases can occur at the interface.
However, the number of interfacial molecules is at least 2 orders of magnitudes less than the number of molecules inside a droplet with a typical size in our experiments.
Although the production rate of hydrogen from the interfacial reaction is expected to be faster thanks to the faster reaction rate, the low number of interfacial molecules at any given time limits the total amount of gas production.
As the diffusion coefficient of hydrogen in organic liquid (droplet phase) is much faster than in water and the small dimension of the droplets (especially in the height), hydrogen produced at the droplets may reach a certain constant concentration in the droplet at equilibrium with hydrogen dissolved in the surrounding in a short time.

In addition to hydrogen from on-droplet production, the in-droplet production of hydrogen analyzed in our model is attributed to water (carrying the catalyst) diffusing into the droplets.
In-droplet water reacts with siloxane to produce hydrogen locally in the droplets.
We note that the presence of co-solvent acetone in the system can increase the solubility of water in the droplets and enhance in-droplet hydrogen concentration from the reaction inside the droplets.
To a certain extent, even for pure siloxane droplets in absence of acetone, it may be impossible to fully eliminate diffusion of water into reactive droplets as the droplet surface is not impermeable.

A final note is that in-droplet reaction may drive continuous intake of water into droplets, in contrast to the equilibrium from the pure partition where the intake of water is limited by the solubility of water in droplet liquid.
The reason is that free water is chemically converted in fast in-droplet reactions, compared to the slow mass transfer of water.
The concentration of water may not reach the solubility limit in the droplet.
The mechanism is similar to what was reported in the latest work on acid-base reactions, and is the reason why the droplet sensing can achieve very high sensitivity \cite{wei2022,hui2022}.

\section{Conclusion}
In summary, the growth rate of hydrogen microbubbles as the gas product was quantified to report the reaction rate inside the microdroplet. The dependence of bubble growth rate on the location inside the droplets, and on the size, morphology and composition of the droplets all suggest that the reaction rate inside microdroplets may be spatially non-uniform. In our theoretical analysis, diffusion of reactant (water) into the droplets is considered to be the rate-limiting step in bubble growth. The predicated local gas concentration is in good agreement with the effects from droplet size, substrate wettability and the composition of binary droplets. 

Understandings of in-droplet bubble formation may lay a foundation for exploring droplet reactions with enhanced chemical kinetics and hydrogen generation. Moreover, nanobubbles encapsulated in surface droplets may lead to a new pathway to functional slippery surfaces to reduce the surface friction \cite{neto2022},  porous surface-bound materials by templating nanobubbles \cite{nanobubble2014}, or coated bubbles used in biomedical imaging, and therapeutic delivery of oxygen or other pharmaceutical compounds\cite{abou2021horizon,fayyaz2021dextran}.

\begin{table}[htbp]
\small
  \caption*{\ List of symbols}
  \begin{tabular*}{1\textwidth}{@{\extracolsep{\fill}}ll}
    \textbf{Symbols} & \textbf{Definition} \\
    $\sigma_{liq-air}$ & Interfacial tension between liquid and air \\
    $\mu$ & Viscosity \\
    $\theta$ & Contact angle of surface microdroplet on the solid substrate \\
    $A_b$ & Base area of surface microdroplet \\
    $a_b$ & Base area of surface microbubble \\
    $t$ & Time calibrated by adding NaOH solution\\
    $R_b$ & Base radius of surface microdroplet \\
    $R_s$ & Radius of the curvature (spherical model)\\
    $r$ & Base radius of surface microbubble \\
    $D_b$ & Distance from bubble center to the droplet rim \\
    $D_s$ & Distance from bubble center to the droplet surface \\
    $\nabla$ & Nabra operator \\
    $D$ & Hydrogen diffusion coefficient in surface droplets \\
    $C_w(D_s)$ & Local water concentration with the distance $D_s$ from the droplet surface \\
    $C_w(D_s)|_{D_s=0}$ & Theoretical water concentration nearing the droplet surface \\
    $C_{SiH}$ & Water concentration in the droplet \\
    $k_r$ & Reaction rate constant \\
    $\epsilon$ & Characteristic diffusion length of water in surface droplets \\
    $\dot m(D_s)$ & Local production rate of hydrogen by the chemical reaction \\
    $\dot m(D_s)|_{D_s=0}$ & Theoretical production rate of hydrogen nearing the droplet surface \\
    $\dot r(D_s)$ & Local average growth rate of hydrogen bubbles\\
    $\dot r(D_s)|_{D_s=0}$ & Theoretical average growth rate of hydrogen bubbles nearing the droplet surface \\
  \end{tabular*}
\end{table}

\begin{acknowledgement}
XHZ acknowledges Future Energy Systems (Canada First Research Excellence Fund), the start-up fund from Faculty of Engineering, University of Alberta, the Canada Research Chairs program, the Canada Foundation for Innovation (CFI) and Alberta-innovates. The work is partly supported by the funding support from the Natural Science and Engineering Research Council of Canada (NSERC).

\end{acknowledgement}

\begin{suppinfo}

\end{suppinfo}

\bibliography{Citations}

\end{document}